\input{aipcheck}

\documentclass[
    ,final            % use final for the camera ready runs
  ]{aipproc}

\layoutstyle{8x11single}
\newcommand{\eqb}{\begin{equation}}
\newcommand{\eqe}{\end{equation}}
\newcommand{\dmb}{\begin{displaymath}}
\newcommand{\dme}{\end{displaymath}}
\newcommand{\pd}{\partial}

\newcommand{\eab}{\begin{eqnarray}}
\newcommand{\eae}{\end{eqnarray}}

\newcommand{\e}{\mbox{e}}
\newcommand{\be}{\begin{equation}}
\newcommand{\ee}{\end{equation}}

\begin{document}

\title{Thermal ground state in Yang-Mills thermodynamics}



\author{Ralf Hofmann}{
  address={ITP, Universit\"at Heidelberg, Philosophenweg 16, 69120 Heidelberg, Germany}
}

\begin{abstract}
We derive an a useful priori estimate for the thermal ground state of deconfining phase of SU(2) Yang-Mills 
thermodynamics in four-dimensional, flat spacetime and discuss 
its implications. Upon a selfconsistent spatial coarse-graining over noninteracting, trivial-holonomy 
(BPS saturated) (anti)calorons of unit topological charge modulus an inert, adjoint scalar field $|\phi|$ and an effective 
pure-gauge configuration $a_\mu^{\mbox\tiny{gs}}$ emerge. The modulus $|\phi|>0$ defines the maximal 
resolution in the coarse-grained theory and induces dynamical 
gauge-symmetry breaking. Thanks to perturbative renormalizability and the fact that 
$|\phi|$ can not absorb or emit energy-momentum the effective 
action is local and simple. The temperature dependence of the effective 
coupling is a consequence of thermodynamical consistency and 
describes the Coulomb screening of a static test charge due 
to short-lived monopole-antimonopole pairs. The latter occur unresolvably as 
small-holonomy excitations of (anti)calorons by the absorption of propagating fundamental gauge fields.      
\end{abstract}

\maketitle

%%%%%%%%%%%%%%%%%%%%%%%%%%%%%%%%%%%%%%%%%%%%
%% MAINMATTER
%%%%%%%%%%%%%%%%%%%%%%%%%%%%%%%%%%%%%%%%%%%%

\section{Introduction}

A purely perturbative approach to high-temperature Yang-Mills thermodynamics is 
inappropriate because the screening effects in the static magnetic sector are too 
weak to inforce the convergence of the expansion after 
resummation of low-order polarization effects \cite{Linde1980}. On the 
other hand, Euclidean 4D SU(2) Yang-Mills theory disposes of topologically 
stabilized, (anti)selfdual field configurations \cite{BPST,ADHM,HarringtonShepard1978,Nahm,LeeLu1998-5,KraanVanBaal1998I-5} 
which, due to a weight $\propto\mbox{e}^{-\frac{8\pi^2 |k|}{g^2}}$ 
($g$ the coupling constant, $k$ the topological charge) in the partition function, 
are ignored by small-coupling expansions. All finite-action (anti)selfdual field configurations 
($F_{4i}=E_i=\pm\frac12\epsilon_{ijk} F_{jk}=\pm B_i\,,\ (i,j,k=1,2,3)$, 
$F_{\mu\nu}=F^a_{\mu\nu}t^a\equiv \partial_\mu A_\nu-\partial_\nu A_\mu-i[A_\mu,A_\nu]$ 
the field strength tensor, $A_\mu=A^a_\mu t^a$ the gauge field, and 
$t^a\,,\ (a=1,2,3)$, generators of SU(2) normalized to tr\,$t^a t^b=\frac12\,\delta^{ab}$) 
share the property that their energy-momentum tensor $\Theta_{\mu\nu}$ vanishes identically:
%%%%%%%%%%%%%%%%%%%%
\begin{eqnarray}
\Theta_{\mu\nu}&=&-2\,\mbox{tr}\Big\{\delta_{\mu\nu}
\left(\mp \vec{E}\cdot\vec{B}\pm\frac14(2\vec{E}\cdot\vec{B}+2\vec{B}\cdot\vec{E})\right)\mp(\delta_{\mu 4}\delta_{\nu i}+\delta_{\mu i}\delta_{\nu 4})\,(\vec{E}\times\vec{E})_i\nonumber\\ 
&&\pm \delta_{\mu i}\delta_{\nu (j\not=i)}\,(E_iB_j-E_iB_j)\pm \delta_{\mu (j\not=i)}\delta_{\nu i}\,
(E_jB_i-E_jB_i)\Big\}\equiv 0\,,.
\end{eqnarray}
%%%%%%%%%%%%%%%%%%%%%%
To exploit this observation is essential in the construction of the thermal 
ground-state estimate in the deconfining phase of the theory: An effective field that is obtained by coarse-graining over noninteracting gauge-field configurations of 
vanishing pressure and energy density can not propagate!

\section{Principles of infinite-volume thermodynamics}

Important constraints in constructing a useful estimate for the thermal 
ground state are imposed by thermodynamics itself. Here we state two basic principles of the 
infinite-volume case. These principles guide the coarse-graining process that involves 
fundamental, action minimizing gauge-field configurations of 
vanishing energy-momentum. They are:\vspace{0.2cm}\\
\noindent I) In the absence of external sources, a thermodynamical gauge system in the 
infinite-volume limit guarantees, up to admissible gauge transformations, 
the spatial isotropy and homogeneity of an {\sl effective} local  
field if this field is {\sl not} associated with the propagation of energy-momentum 
by {\sl fundamental} gauge fields: The partial ensemble average in combination 
with a spatial average, leading to the emergence of such a field, yields a nonzero result only if this field 
is a rotational scalar $\phi$, and $\phi$ does not depend on space 
up to admissible gauge transformations. \\  
\noindent II) Up to admissible gauge transformations the field $\phi$ must not 
depend on time. 
\vspace{0.2cm}\\ 
Principle I) is selfevident, principle II) is a consequence of fact that $\phi$ does not possess energy-momentum at any 
time. 

\section{Calorons}

The only (anti)selfdual or Bogomoln'yi-Prasad-Sommerfield (BPS) saturated fundamental gauge-field 
configurations of SU(2) Yang-Mills theory at high temperature $T$, which 
enter into the coarse-graining process leading to the emergence of the effective scalar field $\phi$, turn out to 
be (anti)calorons \cite{Nahm,HarringtonShepard1978,LeeLu1998-5,KraanVanBaal1998I-5} 
of unit topological charge modulus \cite{HerbstHofmann2004-5,Hofmann2005}. Calorons are instantons of period $\beta\equiv\frac{1}{T}$
in Euclidean time $\tau$. If at spatial infinity the exponential of the integral of $i A_4$ over one period (the Polyakov loop) 
coincides with an element of the center group ${\bf Z}_2$ then the caloron is said to be of trivial holonomy. The trivial-holonomy charge-one caloron, 
which is stable under radiative corrections \cite{Diakonov2004-5} and thus enters into the 
a priori estimate of the thermal ground state in terms of the field $\phi$, was constructed in \cite{HarringtonShepard1978} based on the work 
\cite{'tHooft1976U,JackiwRebbi1976} (particular multiinstantons). In singular gauge, where topological charges are located at instanton centers, 
the trivial-holonomy expressions for a charge-modulus-one caloron (C) or anticaloron (A) read: 
%********
\eqb
\label{gen-inst-sol-log-redgen}
\bar{A}^{C,a}_\mu(x)=-\bar{\eta}^a_{\mu\nu}\,\pd_\nu \log \Pi\,, \ \ \ \bar{A}^{A,a}_\mu(x)=-\eta^a_{\mu\nu}\,\pd_\nu \log \Pi\,,
\eqe
%**********
where 
%********
\eqb
\label{eomWinsolbetaexp}
\Pi(\tau,\vec{x};\rho,\beta)=
1+\frac{\pi\rho^2}{\beta r}
\frac{\sinh\left(\frac{2\pi r}{\beta}\right)}{\cosh\left(\frac{2\pi r}{\beta}\right)-
\cos\left(\frac{2\pi\tau}{\beta}\right)}\,,
\eqe
%*********
$r\equiv|\vec{x}|$, $\eta^a_{\mu\nu}$ ($\bar{\eta}^a_{\mu\nu}$) denotes the selfdual (antiselfdual) 
't Hooft symbol, $\rho>0$ is the size parameter, and the configuration possesses 
no overall magnetic charge. Variablility of $\rho$ 
expresses the invariance of the classical Yang-Mills action tr\,$\frac12\int d\tau d^3x\,F_{\mu\nu}F_{\mu\nu}$ 
under spatial scale transformations. For reasons mentioned below we do 
not make other moduli of these minimal-action configurations explicit. The much harder construction of a nontrivial-holonomy 
charge-one caloron (no overall magnetic charge but magnetic-monopole constituents) is based on the Nahm transformation 
and was carried out explicitly in \cite{LeeLu1998-5,KraanVanBaal1998I-5}. 
While a configuration with static holonomy has vanishing quantum weight in the infinite-volume 
limit \cite{GrossPisarskiYaffe1981} and thus does not contribute to the partition 
function the concept of a temporary holonomy, associated with short-lived magnetic dipoles (small holonomy) or stable, 
screened magnetic (anti)monopoles (large holonomy), nevertheless applies.

\section{Inert, adjoint scalar field}

The (dimensionless) phase $\{\hat{\phi}\}$ of the field $\phi=\{\hat{\phi}\}|\phi|$ ($|\phi|\equiv\sqrt{\frac12\,\mbox{tr}\,\phi^2}$), 
which due to Lie-algebra 
valued gauge fields in the fundamental Yang-Mills action must be an 
adjointly transforming under gauge rotations, is contained in the set defined as  
\eqb
\label{definition}
  \{\hat{\phi}^a\}\equiv\sum_{C,A}\mbox{tr}\,\int d^3x \int d\rho
  \,t^a\,F_{\mu\nu} (\tau,\vec 0) \, \left\{ (\tau,\vec 0),(\tau,{\vec x})
\right\}\,F_{\mu\nu} (\tau,{\vec x}) \, \left\{(\tau,{\vec x}),(\tau,{\vec 0})
\right\} \,,
\eqe
where
\eqb		
\label{abk}
\left\{(\tau,\vec 0),(\tau,\vec x)\right\}\equiv
{\cal P} \exp \left[ i \int_{(\tau,\vec 0)}^{(\tau,\vec x)} dz_{\mu} \, A_{\mu}(z) \right]\,, \ \ \ 
\left\{(\tau,\vec x),(\tau,\vec 0)\right\}\equiv\left\{(\tau,\vec 0),(\tau,\vec x)\right\}^\dagger\,.
\eqe
Since a priori no spatial scale is available on the level of BPS saturation and because of 
spatial isotropy and homogeneity the Wilson lines in Eq.\,(\ref{abk}) need to be computed along 
the straight spatial line connecting the points $(\tau,\vec 0)$ and $(\tau,\vec x)$, and 
${\cal P}$ demands path-ordering symbol. In (\ref{definition}) the sum is over the $|k|=1$ 
caloron ($C$) and anticaloron ($A$) of trivial holonomy. 

It can be shown \cite{HerbstHofmann2004-5,Hofmann2005,Hofmann2007-5} that the definition of the set in 
(\ref{definition}) is unique: no higher nonlocal $n$-point products with factors 
$F_{\mu\nu}$ and no (anti)calorons of higher topological charge modulus may contribute. This set turns out to be 
the kernel of a uniquely defined linear differential 
operator ${\cal D}$. Notice that the kernel and thus the associated operator 
would, due to the (anti)selfduality of the gauge field $A_{\mu}$, be trivial if it were defined in a  
{\sl local} way \cite{HerbstHofmann2004-5}. 

The evaluation of the right-hand side of (\ref{definition}) is quite intricate and 
subtle. In performing it one notices that only the magnetic-magnetic correlation 
contributes in the sense that a divergence of the $r$ integration is multiplied by a 
zero of the angular azimuthal integration to yield an undetermined real scale factor for each polarization 
state in the azimuthal plane. Also, there is a freedom of choice of phase for each polarization. Together this gives a four-dimensional 
real parameter space for oscillatory motion in the plane, two parameters for each oscillation. 
Thus ${\cal D}=\partial^2_\tau+\left(\frac{2\pi}{\beta}\right)^2$. The apparent breaking of rotational 
symmetry introduced by the angular regularization is unmasked as a global choice of 
gauge, there is a fast saturation of the set to the kernel of ${\cal D}$ in the infrared, and no ultraviolet divergence 
arises. The fact that differential operator ${\cal D}$ is linear and that $|\phi|\equiv\sqrt{\frac12\,\mbox{tr}\,\phi^2}$ does not depend on 
space and time implies that ${\cal D}$ annihilates the 
entire field $\phi=|\phi|\hat{\phi}$. By virtue of the Euler-Lagrange equations this in turn 
implies that the field
$\phi$ possesses a canonic kinetic term 
$\mbox{tr}\,(\partial_\tau \phi)^2$ in its effective, Euclidean 
Langrangian density ${\cal
  L}_\phi$. Moreover, the Euler-Lagrange equations for $\phi$ and the (anti)selfduality (BPS saturation) 
of (anti)calorons imply that 
the explicit $\beta$ dependence in ${\cal D}$ be replaced by the 
$\phi$-derivative of a potential $V(\phi^2)$. Therefore 
 \begin{equation}
\label{effactphi}
{\cal
  L}_\phi=\mbox{tr}\left((\partial_\tau\phi)^2+V(\phi^2)\right)\,.
\end{equation} 
We now determine $V(\phi^2)$. The Euler-Lagrange 
equations derivable from Eq.\,(\ref{effactphi}) 
read \cite{GiacosaHofmann2006}
\begin{equation}
\label{ELe}
\partial^2_\tau\phi^a=\frac{\partial V(|\phi|^2)}{\partial
  |\phi|^2}\,\phi^a\ (\mbox{in components})\ \Leftrightarrow\ \partial^2_\tau\phi=\frac{\partial V(\phi^2)}{\partial
  \phi^2}\,\phi\ (\mbox{in matrix form})\,.
\end{equation}  
Since $\phi$'s motion is within a 2D plane in su(2), since $|\phi|$ is independent of
space and time, and since $\phi$'s phase $\hat{\phi}$, viewed as a function of $\tau$, is of 
period $\beta$, one may, without restriction of generality write the solution to Eq.\,(\ref{ELe}) as
\begin{equation}
\label{solEL}
\phi=2\,|\phi|\,t_1\,\exp(\pm\frac{4\pi i}{\beta}t_3\tau)\,.
\end{equation} 
BPS saturation and in particular the vanishing of the Euclidean energy
density together with Eq.\,(\ref{solEL}) imply that 
\begin{equation}
\label{BPSans}
|\phi|^2\left(\frac{2\pi}{\beta}\right)^2-V(|\phi|^2)=0\,.
\end{equation}
Comparing Eq.\,(\ref{ELe}) with $\partial^2_\tau\phi+\left(\frac{2\pi}{\beta}\right)^2\phi=0$, 
we have 
\begin{equation}
\label{ELcomp}
\left(\frac{2\pi}{\beta}\right)^2=-\frac{\partial V(|\phi|^2)}{\partial
  |\phi|^2}\,.
\end{equation}
Together, Eqs.\,(\ref{BPSans}) and (\ref{ELcomp}) yield
\begin{equation}
\label{eomPot}
\frac{\partial V(|\phi|^2)}{\partial
  |\phi|^2}=-\frac{V(|\phi|^2)}{|\phi|^2} \ \ \Rightarrow\ \ \ V(|\phi|^2)=\frac{\Lambda^6}{|\phi|^2}\ \ \Rightarrow \ \ \ 
|\phi|=\sqrt{\frac{\Lambda^3\beta}{2\pi}}\,,
\end{equation}
where the mass scale $\Lambda$ is a constant of integration which can be used to define a 
dimensionless temperature: $\lambda\equiv\frac{2\pi T}{\Lambda}$. In contrast to perturbation theory, where the 
Yang-Mills scale is associated with the Landau pole for the running gauge coupling $g$ dimensional transmutation in the 
effective theory is free of such a contradiction.     

To arrive at the entire effective action valid at maximal resolution $|\phi|$ one exploits that 
perturbative renormalizability \cite{tHooftVeltman1972I-5,tHooftVeltman1972II-5,tHooft1971,LeeZinn-Justin1972} 
implies that coarse-graining out topologically trivial, 
propagating quantum fluctuations down to resolution $|\phi|$ does not alter the form of the action 
density of the associated effective modes. Moreover, gauge invariance demands that 
$\pd_\tau\phi\to D_\mu\phi=\partial_\mu\phi-ie[a_\mu,\phi]$ where $a_\mu$ denotes the 
effective, propagating gauge field, and $e$ is the 
effective coupling constant. Also, the inertness of the field 
$\phi$ due to BPS saturation does not allow for momentum transfer mediated by 
local, higher-dimensional operators, and therefore the effective action density reads    
\begin{equation}
\label{fullactden}
{\cal L}_{\mbox{\tiny eff}}[a_\mu]=\mbox{tr}\,\left(\frac12\,
  G_{\mu\nu}G_{\mu\nu}+(D_\mu\phi)^2+\frac{\Lambda^6}{\phi^2}\right)\,,
\end{equation}
where $G_{\mu\nu}=\partial_\mu a_\nu-\partial_\nu
a_\mu-ie[a_\mu,a_\nu]\equiv G^a_{\mu\nu}\,t_a$ denotes the effective field 
strength, and the full ground-state estimate at tree-level is given 
by $\phi$ and by the pure-gauge configuration 
$a^{\mbox{\tiny gs}}_\mu=\mp\delta_{\mu 4}\frac{2\pi}{e\beta}\,t_3$. The latter solves the Euler-Lagrange 
equations for $a_\mu$ in the effective theory. Notice that $a^{\mbox{\tiny gs}}_\mu$ generates transmutes vanishing pressure and 
energy density of the field $\phi$ into negative and positive values of the ground-state estimate. Microscopically, this relates to an interaction-induced temporary, small 
holonomy for each (anti)caloron implying the creation, subsequent 
collapse, re-creation, ... of monopole-antimonopole pairs. A gauge rotation to unitary 
gauge $\phi=\delta^{3a}\,|\phi|$ reveals the breaking of the electric ${\bf Z}_2$ symmetry associated with a finite expectation of the 
Polyakov loop \cite{Hofmann2005,Hofmann2007-5}. This confirms the deconfining nature of the discussed phase.  

\section{Thermal quasiparticles on tree level}

The effective action density ${\cal L}_{\mbox{\tiny eff}}$ in 
(\ref{fullactden}) has an in-built mechanism for gauge symmetry 
breaking SU(2)$\to$U(1) due to $|\phi|>0$. As a consequence, two out of three 
su(2) directions of propagating gauge modes acquire mass on tree-level. The requirement that this 
mass is thermodynamically consistent, namely, that, at the one-loop level, the energy density of the effective theory can be obtained by 
a Legendre transformation from the effective pressure yields the following evolution equation for 
$e$
\begin{equation}
\label{evalambdasu2}
1=-\frac{24\lambda^3}{(2\pi)^6}\left(\lambda\frac{da}{d\lambda}+a\right)a\,D(2a)\,,
\end{equation} 
where $a\equiv\frac{m}{2T}$, $m=2e\sqrt{\Lambda^3/(2\pi T)}$, and $D(y)\equiv-\frac{4\pi^2}{T^4}\partial_{y^2}
P=\int_0^\infty dx\,\frac{x^2}{\sqrt{x^2+y^2}}\,\frac{1}{\e^{\sqrt{x^2+y^2}}-1}$. There are two fixed points of this evolution: 
One at $\lambda_c\equiv 13.87$, where $e$ exhibits a logarithmic pole, $e(\lambda)\propto -\log(\lambda-\lambda_c)$, 
and one for $\lambda\to\infty$ where $a\to 0$. A plateau value $e=\sqrt{8}\pi$ is reached radidly as $\lambda$ increases away from 
$\lambda_c$. This expresses the fact that the strength of the U(1) Coulomb field of an isolated magnetic test charge is 
renormalized independently of temperature: a nonperturbative manifestation of asymptotic freedom. Effective 
radiative corrections introduce a mass scale into Coulomb-field screening. This is due to the disscociation of (anti)calorons into stable monopole-antimonopole pairs 
through the excitation of large holonomies: The magnetic 
field of dynamically created, stable  (anti)monopoles is short range \cite{Ludescheretal2008}.

\section{Outlook on radiative corrections}

The computation of effective radiative corrections in real-time signature is explained in detail in 
the talks by Markus Schwarz \cite{MS} and Dariush Kaviani \cite{DK}. Here we just spell out the 
main features of the effective loop expansion. In \cite{Hofmann2006} it is conjectured that the expansion into 1-particle 
irreducible diagrams of the polarization tensors of each gauge mode (reducible diagrams are resummed) 
terminates at finite orders. This carries over to the loop expansion of any thermodynamocal quantity, say the pressure. 
The argument uses the fact that nonempty intersections of the 
algebraic varieties defined by the constraints on the loop variables imposed by the 
maximal resolution $|\phi|$ become extremely unlikely at a sufficiently large number of (independent) loops. It is these intersections, however,  
that support the loop intergrations. To make this plausible, the number of constraints 
on a priori noncompact loop variables 
as a function of loop number can be estimated by appealing to the 
Euler- L' Huilliers characteristics of the polyhedron associated with a 
given loop diagram. One observes that the number of constraints starts to exceed the 
number of loop variables very early in the loop expansion. Numerically, we investigated the two- and 
three-loop situation in the pressure expansion, and we see very fast convergence.      

\section{Summary}

This talk is on the construction of a thermal ground-state estimate for deconfining SU(2) 
Yang-Mills thermodynamics. In a first step, 
this construction invokes a spatial coarse-graining over noninteracting (anti)calorons of unit charge modulus 
and trivial holonomy. The associated emergence of an inert, adjoint scalar field $\phi$ and perturbative renormalizability 
imply a simple, local effective action valid for a maximal resolution $|\phi|$. We derive a tree-level ground-state estimate for this effective theory and 
discuss its implications for the spectrum and the evolution of the effective coupling. Finally, 
we give a brief outlook on effective radiative corrections.

\section*{Acknowledgments}

\noindent I would like to express my gratitude to the following Diploma students at the universities of Heidelberg and Karlsruhe and the 
Karlsruhe Institute of Technology who, over a period of seven years, helped to develop the ideas and results discussed in this symposium: 
Ulrich Herbst, Jochen Rohrer, Markus Schwarz, Dariush Kaviani, Michal Szopa, Jochen Keller, Sebastian Scheffler, Julian Moosmann, Josef Ludescher, and 
Carlos Falquez. To Francesco Giacosa, who, in an exemplary way and under difficult conditions, committed himself to common research and 
to sober, constructive criticism, go my very special thanks. I feel indebted to Theodore Simos for his suggestion 
to organize this meeting and for his outstanding efforts in running the International Conferences on 
Numerical Analysis and Applied Mathematics.

%%%%%%%%%%%%%%%%%%%%%%%%%%%%%%%%%%%%%%%%%%%
%% The following lines show an example how to produce a bibliography
%% without the help of the BibTeX program. This could be used instead
%% of the above.
%%%%%%%%%%%%%%%%%%%%%%%%%%%%%%%%%%%%%%%%%%%

\end{document}